\documentstyle[12pt]{article}
\begin{document}
\def\BB#1{{\Large \bf #1}}
\pagestyle{empty}
\begin{center}
\hfill MPI--Ph/92--03 \\
\hfill August 1992\\[2.15cm]
{\Large \bf \BB{L}eading \BB{H}iggs \BB{M}ass \BB{D}ependent
\BB{C}orrections \\to the
\BB{S}tandard \BB{M}odel \BB{L}agrangian\\[0.1in]
and its \BB{T}wo \BB{H}iggs \BB{D}oublet \BB{E}xtension
} \end{center}
\vskip 1.2cm
\begin{center} {\large Vidyut Jain\footnote{Current Address:
Theoretical Physics
Group, 50A-3115, L.B.L., Berkeley, CA 94720.} } \\[0.13cm]
{\it Max-Planck-Institut f\"ur Physik,\\ F\"ohringer Ring 6,
 D-8000 M\"unchen 40, Germany }\end{center}
\vskip 1.5cm
\begin{center} {\large \bf Abstract} \end{center}
We present an alternative calculation for the leading
Higgs mass dependent one--loop corrections to the standard model
Lagrangian, using a background field technique. Cross--sections
computed from our one--loop Lagrangian provide a check of and
reproduce results already obtained in the existing literature
using diagrammatic methods, as well as allowing an analysis of
other processes involving longitudinally polarized $W^\pm$s and
$Z^0$s. We concentrate on the processes $WW\rightarrow f\bar{f}$
and $f\bar{f}\rightarrow WW$. We extend our result to the case
of a two Higgs doublet model, when one of the physical Higgs is
much heavier than the other particles and analyze the tree+one--loop
unitarity bounds on the scale at which signs of a strongly
interacting sector (or the heaviest Higgs) must turn up.

\newpage
\def\Tr{{\textstyle Tr}}
\def\tr{{\textstyle tr}}
\def\({\left (}
\def\){\right )}
\def\[{\left [}
\def\]{\right ]}
\pagestyle{plain}
\setcounter{page}{1}
\noindent {\large \bf 1.  Introduction.}

It has become clear that the standard model of gauge interactions
has yielded a remarkably good description of nature. In spite
of this, we remain completely ignorant of the symmetry breaking
sector: we do not know if it is described by a simple Higgs
structure as in the minimal standard model; we do not know if this
sector is described by elementary scalars, or composite ones such
as in technicolor scenarios; we do not know if its interactions
are strong or weak.

General considerations [1] lead one to expect that something
interesting will turn up at the next round of colliders (with
the possible exception of the ``intermediate mass" Higgs).
Unitarity of partial wave amplitudes implies that something
will turn up by about 1 to 2 TeV, for example a Higgs particle or
signs of a strongly interacting symmetry breaking sector. In this case,
even if no physical resonance is detected at, e.g., the LHC or SSC,
it will still be possible to learn something about the symmetry
breaking sector by studying processes involving longitudinally
polarized $W$s and $Z^0$s since these are the components that
arise from symmetry breaking.

In this paper we present the one--loop corrections to the standard
model Lagrangian that grow as $M_H^2$ or $\ln M_H$. We include
all loops containing electroweak gauge bosons, pseudoscalars, and
fermions. We used a background field technique which yields a
manifestly gauge invariant tree+one--loop Lagrangian, and work
within the context of a nonlinear $\sigma$--model. Our calculation
constitutes the evaluation of many Feynman diagrams and provides
a very important independent check of various partial results already
found in the existing literature. With our Lagrangian it is also
possible to extract easily other one--loop corrections.
Additionally, we extend our results to include the two Higgs doublet
model when one of the neutral physical Higgs is much heavier than the
other particles.

The current paper is the first step in finding the leading Higgs
mass and top mass dependent corrections to the standard model, and
its two doublet extension. For the case of the standard model,
leading fermion mass dependent corrections to the tree level Lagrangian
have been computed by the author in [2]. However, the physical
relevance of these corrections were not discussed there.
Their
significance to physics at colliders will be presented in a
forthcoming paper. Together with the results of this paper, they
will represent the leading corrections to many interesting
physical processes.

There is already an extensive amount of literature on leading
Higgs mass dependent one--loop corrections, using both diagrammatic
[3,4] techniques and the background field method [5,6,7].
In ref. [6], the
background field method was used to find the one loop corrections
to the bosonic part of the standard model
Lagrangian from loops containing bosons only, and
in ref. [7] this was extended to include external fermions.
Our calculation gives the complete result by including fermions
as both external particles and internal particles. Our additional
corrections to the Lagrangian in the case of an extra Higgs doublet
are also new. We have checked our results against these previous
background field calculations. Furthermore, for loops containing
fermions, we have checked our results against the complex diagrammatic
calculations of ref. [8] for the leading corrections to the process
$f \bar{f}\rightarrow W_L W_L$.

Apart from the tree+one--loop Lagrangian, the main new results
of this paper are summarized.
\begin{itemize}
\item For a top quark mass closer to the CDF limit [9] than the
 perturbative upper bound [10] we are in agreement with ref. [8]
 and find a 15\% enhancement over the standard model
tree amplitude for the
 process $\bar{t}t\rightarrow W^+_L W^-_L$ for $M_H\approx 2$TeV
and $\sqrt{\hat{s}}\approx 1$TeV. This may be modified for higher
values of the top quark mass.
In addition, ref. [7] gives a partial analysis of heavy pair production
 from $W_L^+ W_L^- \rightarrow \bar{f} f$ in the standard model.
With our full Lagrangian we find no significant one--loop correction
to such a process.
\item In the two Higgs doublet model we find corrections to
 processes involving not only longitudinally polarized gauge bosons,
but also the $CP$--odd Higgs and the light $CP$--even Higgs.
It is known that tree-level unitarity limits on the scale of
new physics in a two Higgs doublet model with a very heavy
Higgs can be significantly stronger than in the minimal standard
model [11]. We analyze similar unitarity limits for tree+one--loop
amplitudes and find that the scale at which new physics must
enter to unitarize the amplitudes to be in general lower than at
tree level. In fact, for fixed $M_H$, we find this scale to fall
extremely rapidly with $\tan\beta$, the ratio of the $vev$s.
\end{itemize}

Our analysis is based on the following observation [12]. In the standard
model the purely scalar sector of the Lagrangian is
\begin{equation}
 L = {1\over4} v^2 Tr\[\partial_\mu\Phi^{\dag}\partial^\mu\Phi\]
  -{\lambda\over4}v^4(Tr[\Phi^{\dag}\Phi]/2-1)^2,
\end{equation}
where $v$ is the Higgs $vev$, $\lambda$ is the self--scalar coupling,
$\Phi$ parameterizes the four scalars $\sigma$ (the physical Higgs),
$\pi^1,\pi^2,\pi^3$:
\def\te{\mbox{\boldmath $\tau$}}
\def\pe{\mbox{\boldmath $\pi$}}
\def\ze{\mbox{\boldmath $\zeta$}}
\def\xe{\mbox{\boldmath $\xi$}}
\def\Re{{\bf Re}}
\def\Im{{\bf Im}}
\begin{equation}
  \Phi = \sigma+2i\te\cdot\pe ,
\end{equation}
normalized so that the scalar fields are dimensionless.
Here $\te$ are the three pauli matrices normalized so that $\te\times
\te=i\te$, and boldface denotes 3--vectors.
 We can also arrange these fields in the usual complex
Higgs doublet:
\begin{equation}
 \phi =v\( \begin{array}{c} i\pi^1+\pi^2 \\ \sigma-i\pi^3 \end{array}
   \).
\end{equation}

The perturbative Higgs mass, $M_H=350\sqrt{\lambda}$GeV,
increases with $\lambda$. For $\lambda$ bigger than
$\approx 10$ perturbation theory
breaks down and the quantity $M_H$ no longer corresponds to the
mass of any physical resonance and it becomes difficult to perform
reliable calculations. However, one expects the appearance of a
spectrum of new resonances above some energy.

One approach to calculating loop corrections in such a scenario [12]
is to formally take the limit $\lambda\rightarrow\infty$. The
potential in (1) is proportional to $(\sigma^2+\pe^2-1)^2$ so that
$\lambda\rightarrow\infty$ requires $\sigma\rightarrow\sqrt{1-\pe^2}$
for the energy density to remain finite. The physical Higgs is
removed from the theory in this limit, and the effective Lagrangian
describing the $\pe$ fields is given by the nonlinear $\sigma$--model
\begin{equation}
 L = {1\over2} v^2 g_{ij}\partial_\mu\pi^i\partial^\mu\pi^j,
\end{equation}
with the metric on the space of scalars is given by
\begin{equation} g_{ij}=\delta_{ij}+{\pi^i\pi^j\over 1-\pe^2} .
\end{equation}
The inverse is $g^{ij}=\delta^{ij}-\pi^i\pi^j$.

To describe the standard model with large $\lambda$ one has also
to add gauge fields and fermions. In any case, because (4) describes
a nonrenormalizable model, so will the corresponding effective
Lagrangian for the standard model. In spite of this, loop calculations
performed with it can be used to extract meaningful results. A classic
example of this is the Fermi theory of weak interactions which is
neither unitary nor renormalizable, yet yields a good description
of four fermion interactions below the $W$--boson mass. Being
nonrenormalizable, loop corrections in the Fermi theory contain
divergent terms not in the original Lagrangian. However, since the
Fermi theory can be viewed as a low energy limit of the standard
model (i.e. $M_W\rightarrow \infty$), we may expect that loop
corrections computed with the Fermi theory should have some
correspondence to loop corrections computed with the renormalizable
theory. Indeed, it can be shown [12] that by interpreting the regulating
scale in the Fermi case as of order the $W$--boson mass (i.e. the
scale at which the $W$--boson enters to dampen the otherwise divergent
integrals) one can reproduce surprisingly well many of the corresponding
standard model calculations. In some instances, however, equating the
regulating scale with the $W$--boson mass may very much overestimate
certain processes which in the standard model are suppressed by other
means, for example the GIM mechanism. Generally speaking then, one
should take the regulating scale as the scale at which new physics
not described by the Fermi theory becomes important. Indeed, in the
case of strangeness--changing neutral current transitions, calculations
of this kind were used to predict the charm quark mass [13] even before
the underlying theory was known. [Of course, there are some
renormalizable
divergent corrections which are identical to those found in the full
underlying theory, such as the corrections which give the running of
the gauge couplings.]

We will therefore work within the context of a nonrenormalizable
$\sigma$--model and interpret the regulating (cutoff) scale $\mu$ for the
divergent integrals as the scale at which new physics enters. We
emphasize two possibilities.
\begin{itemize}
\item The symmetry breaking sector is  strongly
interacting.
Then $\mu$ is the scale of the lightest physical resonance
associated with the scalar sector, generally believed to be in the
$1$TeV to 2TeV range [1].
\item The standard model with elementary scalars is not strongly
interacting. It has been shown that in the gauged version of (4) that
if one takes $\mu$ as the perturbative Higgs mass $M_H$ then one
reproduces the leading Higgs mass dependent one--loop corrections
(i.e. those that grow with $M_H^2$ or $\ln M_H$) as computed in
the full standard model [5,6]. In fact, with the addition of fermions
this remains true, and our results bear this out when we can compare
with previous calculations.
\end{itemize}

\noindent{\bf The Lagrangian with one Higgs doublet.}

The construction of the phenomenological Lagrangian describing the
standard model in the limit of infinite self--scalar coupling can be
found in [2]. The Lagrangian is
\def\FF{{\bf F}}
\def\DS{D\!\!\!\!/}
\def\hf{{\textstyle {1\over2}}}
\begin{eqnarray}
L_{SM}&=& -{1\over4g^2}\FF_{\mu\nu}\cdot\FF^{\mu\nu}-{1\over4 g'^2}
G_{\mu\nu}G^{\mu\nu}+{1\over2}v^2 g_{ij} D_\mu\pi^i D^\mu\pi^j
+i\bar{\psi}^a_L\DS\psi_L^a+i\bar{\psi}^a_R\DS\psi_R^a \nonumber \\& &
-v(\lambda_u^{ab}\bar{q}^a_L\Phi[\hf+\tau_3]q_R^b+
  \lambda^{ab}_d\bar{q}^a_L\Phi[\hf-\tau_3]q^b_R+
  \lambda_e^{ab}\bar{l}^a_L\Phi[\hf-\tau_3]l^b_R+H.c.).
\end{eqnarray}
The $SU(2)_L$ field strength $\FF_{\mu\nu}$ and $U(1)_Y$ field
strength $G_{\mu\nu}$ are given by
\def\AA{{\bf A}}
\begin{eqnarray}
 \FF_{\mu\nu} &=& \partial_\mu \AA_\nu-\partial_\nu\AA_\mu +
 \AA_\mu\times \AA_\nu, \nonumber \\
 G_{\mu\nu} &=& \partial_\mu B_\nu-\partial_\nu B_\mu,
\end{eqnarray}
where $\AA_\mu$ are the three $SU(2)_L$ gauge fields and $B_\mu$ is
the $U(1)_Y$ gauge field. The gauge covariant derivative on the
scalars is
\begin{equation}
  D_\mu\pi^i=\partial_\mu\pi^i+\hf \epsilon_{\;\;jk}^i A_\mu^j\pi^k
      -\hf A_\mu^i\sqrt{1-\pe^2} +\hf\epsilon^i_{\;\;3k} B_\mu\pi^k
      +\hf \delta^i_3 B_\mu \sqrt{1-\pe^2}.
\end{equation}
The standard model has left--handed fermion doublets transforming
like $\phi$ under $SU(2)_L$, and with different $U(1)_Y$ charges.
The covariant derivative on left--handed quarks and leptons  is
\begin{eqnarray}
  D_\mu q_L^a &=& (\partial_\mu-i\AA_\mu\cdot\te-{i\over6} B_\mu
     -{i\over2 }C_\mu^i\lambda_i) q_L^a ,\nonumber \\
  D_\mu l_L^a &=& (\partial_\mu-i\AA_\mu\cdot\te+{i\over2} B_\mu)
l_L^a.
\end{eqnarray}
Here, and in (6), the superscripts on the fermions label generation
($a=1\ldots 3$). For the right--handed fermions,
\begin{eqnarray}
  D_\mu u_R^a &=& (\partial_\mu-{2i\over3} B_\mu
     -{i\over2 }C_\mu^i\lambda_i) u_R^a ,\nonumber \\
  D_\mu d_R^a &=& (\partial_\mu+{i\over3} B_\mu
     -{i\over2 }C_\mu^i\lambda_i) d_R^a ,\nonumber \\
  D_\mu e_R^a &=& (\partial_\mu + i B_\mu) e_R^a.
\end{eqnarray}
In (6), we have not written the fermion kinetic terms explicitly;
$\psi$ stands for all the quark and lepton fields. The second line
of (6) gives the Yukawa couplings (normally these are given
in terms of the Higgs fields $\phi$ and $2i\tau_2\phi^*$).
$\lambda^{ab}_u$, $\lambda^{ab}_d$, and $\lambda^{ab}_e$ are
numerical matrices proportional to the usual Yukawa couplings,
and for convenience we have defined $l_R^a=(0,e_R^a)$ and
$q_R^a=(u_R^a, d_R^a)$.

Notice that we have normalized the gauge fields so that the gauge
coupling constants $g$, $g'$ (for $SU(2)$, $U(1)$) appear only as
overall factors in (6). Additionally, we have not kept the kinetic
term for the gluons as we will not calculate loop corrections due
to them.

\noindent{\bf The Lagrangian with two Higgs doublets.}

There is a simple extension of the standard model which is of
current phenomenological interest. We add one more Higgs doublet $\chi$
to the model. Under general considerations [14] the potential
is taken to be
\def\pp{\phi^{\dag}\phi-v^2}
\def\cc{\chi^{\dag}\chi-v'^2}
\def\ppp{\phi^{\dag}\phi}
\def\ccc{\chi^{\dag}\chi}
\begin{eqnarray}
  V(\phi,\chi) &=& \lambda_1(\pp)^2+\lambda_2(\cc)^2\nonumber \\
   &&+ \lambda_3 [(\pp)+(\cc)]^2 +\lambda_4 [(\ppp)(\ccc)
     - (\phi^{\dag}\chi)(\chi^{\dag}\phi)] \nonumber \\
   &&+ \lambda_5 [\Re(\phi^{\dag}\chi)-vv' \cos\xi]^2
     + \lambda_6 [\Im(\phi^{\dag}\chi)-vv' \sin\xi ]^2,
\end{eqnarray}
where $v$ and $v'$ are the two $vev$s that break $SU(2)_L\times U(1)_Y$
down to $U(1)_{EM}$.The $\chi$ doublet is parameterized in terms of
four real fields:
\begin{equation}
 \chi =v'\(\begin{array}{c} i\zeta^1+\zeta^2 \\
     \zeta^0-i\zeta^3 \end{array}
   \).
\end{equation}
An important parameter that we will use later is $\tan\beta=v'/v$.
The properly normalized
scalars that are ``eaten'' by the gauge fields are no longer
purely $\pe$ as in the one doublet case, but
\begin{equation}
   w^i = v\pi^i\cos\beta + v'\zeta^i\sin\beta,
\end{equation}
while the properly normalized $CP$--odd physical Higgs are
\begin{eqnarray}
  H^\pm &=& - v\pi^\pm\sin\beta +v' \zeta^\pm \cos\beta, \nonumber \\
  A^0 &=& -\sqrt{2}v\sin\beta\pi^0 + \sqrt{2}v'\zeta^3 \cos\beta,
\end{eqnarray}
where $\pi^\pm = {1\over\sqrt{2}}(\pi^i\mp i\pi^2)$, etc.

The limit of large $\lambda_1$ corresponds to taking the (perturbative)
mass of one of the $CP$--even
Higgs scalars as large. In fact, in this limit the $CP$--even
physical mass eigenstates are [14]
\begin{eqnarray}
    H^0 &=&\sqrt{2} v (\sigma- 1),
        \;\;\;\qquad\qquad M_{H^0}^2=2v^2\lambda_1
      ,\nonumber \\
    h^0 &=&\sqrt{2} v' (\zeta^0 -1), \qquad\qquad M_{h^0}^2=2v'^2
     (\lambda_2+\lambda_3)+{1\over4}v^2\lambda_5.
\end{eqnarray}
As with the one Higgs doublet model, when we
take $\lambda_1$ to infinity, finiteness of the potential
energy imposes the constraint $\ppp=v^2$, or $\sigma=\sqrt{1-\pe^2}$,
and
$H^0$ is removed from the theory. This is the extension of the standard
model that we include here. Note, that at tree level the masses
of the physical charged Higgs states and the $CP$--odd physical Higgs
state depend on $\lambda_4$ and $\lambda_6$, respectively, and are
not affected by this limit.

The $roles$ of $\phi$ and $\chi$ in the potential (11) are symmetric,
even if the equation is not.
Taking $\lambda_2$ to infinity instead of $\lambda_1$
reverses the role of $\phi$ and $\chi$.
We can recover this case     by swapping
some of the parameters in (11). We do not consider the limit
$\lambda_1\rightarrow\infty$, $\lambda_2\rightarrow\infty$ which
corresponds to taking the perturbative masses of both $CP$--even
Higgs to infinity. Nor do we consider the limit $\lambda_3\rightarrow
\infty$, which corresponds to taking the perturbative mass of
one of the the $CP$--even Higgs eigenstates to infinity and the other
mass to zero. The constraint equation in this last case does not
lead to a scalar kinetic energy of the form (4).

In the limit $\lambda_1\rightarrow\infty$, the scalar potential
becomes
\begin{eqnarray}
  V &=& {1\over4} \lambda'(\cc)^2
    +\lambda_4 [(\ppp)(\ccc)
     - (\phi^{\dag}\chi)(\chi^{\dag}\phi)] \nonumber \\
   &&+ \lambda_5 [\Re(\phi^{\dag}\chi)-vv' \cos\xi]^2
     + \lambda_6 [\Im(\phi^{\dag}\chi)-vv' \sin\xi ]^2,
\end{eqnarray}
where $\lambda'\equiv 4\lambda_2+4\lambda_3$.
The complete Lagrangian is the sum of three terms
\begin{equation}
  L_{SME} = L_{SM} - V + L',
\end{equation}
where $L'$ contains the gauge covariant
kinetic term for the additional Higgs
doublet and its coupling to the fermions:
\begin{eqnarray}
  L' &=&  (d_\mu \chi)^i (d^\mu \bar{\chi})_i\nonumber \\ & &
-v'(\lambda_u^{'ab}\bar{q}^a_L\Xi[\hf+\tau_3]q_R^b+
  \lambda^{'ab}_d\bar{q}^a_L\Xi[\hf-\tau_3]q^b_R+
  \lambda_e^{'ab}\bar{l}^a_L\Xi[\hf-\tau_3]l^b_R+H.c.).
\end{eqnarray}
Here $i=1,2$.
The covariant derivative on $\chi$ is
\begin{equation}
  d_\mu\chi = (\partial_\mu-i\te\cdot\AA_\mu-{i\over2}B_\mu)\chi.
\end{equation}
We have also defined
\begin{equation}
   \Xi = \zeta^0 + 2 i \te\cdot\ze .
\end{equation}
The last line in (18) describes the Yukawa couplings of the $\chi$
fields to the fermions. To suppress tree level flavour changing
neutral currents induced by scalar exchange one normally
concentrates on specific models, for example case (Ia)
the choice that all the $\lambda_u,
\lambda_d,\lambda_e$ are zero while none of the $\lambda'_u,
\lambda'_d,\lambda'_e$ are zero, case (Ib) the opposite situation,
case (IIa) when only $\lambda_d,\lambda_e,\lambda'_u$ are nonzero,
and case (IIb) when only $\lambda'_d,\lambda'_e,\lambda_u$ are
zero. Models IIa and IIb correspond to the Higgs coupling
structure in the minimal supersymmetric standard model, but with
the roles of $\phi$ and $\chi$ interchanged. Our calculations will
be general, but these cases are phenomenologically interesting.

In Section 2 we give the tree+one--loop effective Lagrangian for
both the one Higgs doublet case and the two Higgs doublet case.
In Section 3 we discuss physical applications,
and a reader who is only interested
in physical results can skip Section 2.

\newpage
\noindent{\large\bf 2. One--Loop Corrections.}

In this section we present the leading Higgs mass dependent one--loop
corrections to the one Higgs doublet Lagrangian, eq. (6), and the
two Higgs doublet Lagrangian, eq. (17). We found the one--loop
corrections by using the background field expansion method together
with some functional methods to evaluate one--loop determinants. In this
approach, one splits the fields into background and quantum parts
$\pe=\tilde{\pe}+\hat{\pe}$,
$\AA_\mu=\tilde{\AA}_\mu+\hat{\AA}$, $B_\mu=\tilde{B}_\mu+
\hat{B}_\mu$, $\psi=\tilde{\psi}+\hat{\psi}$,and $\chi=\tilde{\chi}
+\hat{\chi}$,
and expand to second order in the quantum (hatted) fields.
(One can also
keep gluon backgrounds $\tilde{C}^i_\mu$, but we do not).  The
terms linear in the quantum fields, along with appropriate source terms,
are set to zero by the classical equations of motion satisfied by the
background fields. One then functionally integrates over the
quantum quantities in the path integral to find the one--loop effective
action as a function of the background fields. If carefully done
it allows one to keep manifest all the tree level symmetries of the
theory in terms of the background fields.

However, using the fields $\hat{\pe}$ as our quantum fields is not the
best way to proceed. The Lagrangian (6) or (17) is manifestly invariant
under reparameterizations of the scalar fields, i.e. under
$\pi^i\rightarrow \Pi^i(\pe)$, since $\partial_\mu\pi^i$ transforms
as a vector with the scalar metric $g_{ij}$ transforming as a
tensor. Due to the geometrical nature of the covariant derivative
(8) the gauged nonlinear $\sigma$--model also has this property.
The coordinate $\hat{\pi}^i$ does not transform as a vector under
reparameterizations so that an expansion of the action -- a scalar
-- as a power series in the $\hat{\pe}$ is not manifestly invariant
under reparameterizations. Functionally integrating over the
$\hat{\pe}$
then leads to an effective
action which is not manifestly invariant under
(background field) reparameterizations. Although we cannot expect this
to affect physical results, the extraction of such results is easiest
done in a framework in which the reparameterization invariance is
kept manifest.

Such a framework is well known [15]. Instead of $\hat{\pe}$, a
reprametrization vector $\xe$ is chosen as the quantum field to
be integrated over in the path integral. Mukhi [15] has given a very
simple algorithm to generate the expansion of the action
$S[\pe]$ in terms of the three $\xi^i$:
\begin{eqnarray}
 S[\tilde{\pe},\xe]=\sum_{n=0}^\infty {1\over n!} \[
   \int d^4 x \xe(x) \cdot {\bf D}^x\]^n S[\tilde{\pe}],
\end{eqnarray}
where $D^x_i$ is the covariant functional derivative (with respect to
$\tilde{\pi}^i$). For example, on quantities which are scalars under
reparameterizations (such as the action) it is just the ordinary
functional derivative, while on reparameterization vectors it is
constructed with the use of the ``scalar connection" which depends
on the scalar metric $g_{ij}$. One can give explicit formulas but
we will not need them here.

With the background expansion complete we
can use the following result [5,16,2].
 For quantum bosonic fields $\theta$ (a column vector),
and four component spin--${1\over2}$
fermions $\hat{\psi}^A$, the Lagrangian quadratic in the quantum fields
may be generally written as
\begin{equation}
  L = -{1\over2}\theta^T Z[d_\mu
d^\mu +M^2_\theta]\theta + \hat{\bar{\psi}}
  \Delta^{-1}_\psi\hat{\psi}+\[(\theta^T\bar{\Sigma})_A\hat{\psi}^A
  + H.c.\] + L_{G.F.}.
\end{equation}
Here $Z$ is a matrix which gives the bosonic metric, an
extension of (5);
$d_\mu$ is a covariant derivative (gauge, reperameterization, etc.);
$\Sigma$ is a matrix
with mixed bosonic and fermionic indices which describes
the mixing between
bosons and fermions; $M_\theta^2$ is the mass squared
matrix for the bosons; $L_{G.F.}$ includes the terms that arise from
gauge fixing (auxiliary fields and ghosts) all the local invariances;
$\Delta^{-1}_\psi=i\DS_R R+ i\DS_L L - M_\psi$, where $D_R^\mu$ is the
covariant derivative on right--handed fermions ($D_L^\mu$ is the
covariant derivative
on left--handed fermions), $R$ ($L$) is the projection
operator for right--handed (left--handed) fermions, and $M_\psi$ is
the mass matrix for the
fermions which can be decomposed into right-- and
left-- handed parts, $M_\psi= m^{\dag} R + m L$.

The divergent (regulated) one--loop corrections from just the bosonic
fields is [5]
\begin{equation}
\delta L_\theta=
-{1\over 2^6\pi^2}\Tr \[(M_\theta^4+{1\over6}J^{\mu\nu})
\ln [2\mu_0^2/\mu^2]+4M_\theta^2\mu^2\ln 2\].
\end{equation}
[We have dropped total divergences, as we do throughout.]
Here $\mu$ is the regulating scale, $\mu_0$ is a scale
characterizing the low energy theory and $J_{\mu\nu}=[d_\mu,d_\nu]$.
When $m$ does not contain any $\gamma$--matrices, the divergent
one--loop corrections from just the fermions is [5,16]
\begin{eqnarray}
 \delta L_\psi &=& {1\over 2^4\pi^2}\Tr \[ \left\{
  (m^{\dag}m)^2-D_\mu m D^\mu m^{\dag} \right. \right.\nonumber \\
  & & \qquad\qquad \left.\left. -{1\over6}(F_{\mu\nu}^R F^{R\mu\nu} +
     F^L_{\mu\nu} F^{L\mu\nu})\right\}\ln[2\mu^2_0/\mu^2]
   + 4m^{\dag}m\mu^2\ln 2 \] .
\end{eqnarray}
The trace is over internal indices, the covariant derivative on the
masses is
\begin{eqnarray}
  (D_\mu m^{\dag}) &=& D^L_\mu m^{\dag}-m^{\dag}D_\mu^R, \nonumber \\
  (D_\mu m) &=& D^R_\mu m-m D_\mu^L,
\end{eqnarray}
and $F_{\mu\nu}^{R,L}$ are the Yang--Mills field strengths
$F^R_{\mu\nu}=[D^R_\mu, D^R_\nu]$ and $F^L_{\mu\nu}=[D^L_\mu,D^L_\nu]$.
In the last equation, the derivatives act on everything to the right.
For the $\Sigma$ contributions and the gauge--fixing Lagrangian
we have the corrections [2]
\begin{eqnarray}
 \delta L' &=& {1\over 2^6\pi^2}\ln[2\mu_0^2/\mu^2]
    \Tr\[4 Z^{-1}\bar{\Sigma}M_\psi^{\dag}\Sigma-
        2Z^{-1}\bar{\Sigma}(i\gamma^\mu d_\mu\Sigma)\] \nonumber\\
 & & +{1\over2^5\pi^2}\Tr\[2\mu^2\tilde{M}^2\ln 2 + \left\{
     {1\over2}\tilde{M}^4-{1\over4}S^2\right\}\ln[2\mu_0^2/\mu^2]\]
    \nonumber \\
& & + {1\over2^6\pi^2}\ln[2\mu^2_0/\mu^2] \Tr g^2
\[\bar{\Sigma}M_\psi^{\dag}
       \Sigma-\bar{\Sigma}(i\gamma^\mu d_\mu\Sigma) \right.
  \nonumber \\ & & \qquad\qquad\qquad\qquad\left.
 +{1\over3}(\bar{\Sigma}i\gamma^\nu d^\mu \Sigma)^{\alpha\beta}
    (\eta_{\alpha\beta}\eta_{\mu\nu}+\eta_{\alpha\nu}\eta_{\beta\nu}
      +\eta_{\alpha\mu}\eta_{\beta\nu})\] \nonumber \\
   & & + {2\over 2^6\pi^2}\ln[2\mu^2_0/\mu^2]\Tr[{1\over6}
             J^A_{\mu\nu} J^{\mu\nu}_A ] .
\end{eqnarray}
In the first line, the trace is over all bosonic indices, whilst
for the remainder the trace is only over gauge indices. $\tilde{M}^2$
is the (background) mass squared term for the gauge fields, and $S$
describes the mixing between the gauge fields and the other scalar
fields. More precisely if $\theta^T=(\hat{A}^\mu_I, \hat{\phi}^i)$
then the Lagrangian contains the terms $L\ni \hf \hat{A}^\mu_I
 \eta_{\mu\nu}(\tilde{M}^2)^{IJ}\hat{A}^\nu_J+\hat{\phi}^i
  (S_\mu)^I_i\hat{A}^\mu_I$. Furthermore $d_\mu\Sigma$ is the
covariant derivative acting on $\Sigma$, i.e. that it transforms
in the same way as $\Sigma$ under some (background) transformations.
Finally, $g^2$ in (26) is a diagonal matrix in gauge space with
entries $(g^2,g^2,g^2,g'^2)$.

With these rather general results it is straightforward to
complete the background field expansion and computing the explicit
corrections given by equations (23), (24), and (26).
For the Lagrangian (6) the details of the calculation are given
in [2]. In what follows we dropped the tildes denoting background
fields.

\noindent {\bf One Higgs Doublet Model.}

The complete leading tree+one--loop Lagrangian from (6) is:
\begin{eqnarray}
 L_{SM}(\mu^2)&=& -{1\over4 g^2} Z_g
    \FF_{\mu\nu}\cdot\FF^{\mu\nu}- {1\over 4g'^2}
 Z_{g'} G_{\mu\nu}G^{\mu\nu}
 +{1\over2}v^2 Z_v g_{ij}D_\mu\pi^i D^\mu \pi^j
    \nonumber \\ & &
 +i\bar{\psi}_L Z_L\DS\psi_L + i\bar{\psi}_R Z_R\DS \psi_R
  -v[\bar{\psi}_L\Phi Z_\Lambda \Lambda\psi_R+H.c.]
   \nonumber \\ & &
  +{3\eta\mu^2\over 32\pi^2} g_{ij} D^\mu\pi^i b_\mu^j \sigma
   + {3\ln[\mu_0^2/\rho \mu^2]\over 128\pi^2} v^2 (g^2
      S^\mu_{ip} g^{ij} S_{\mu\; jq}\delta^{pq}+g'^2
      S^\mu_{i0} g^{ij} S_{\mu\; j0})
    \nonumber \\ & &
 -{\ln[\mu_0^2/\rho\mu^2]\over 64\pi^2}\left \{ {2\over3}
   (g_{ij}D^\mu\pi^i D_\mu\pi^j)^2 + {4\over 3}(g_{ij} D_\mu
    \pi^i D_\nu\pi^j)^2 \right.\nonumber \\& &
   \qquad +3\( g_{ij}D^\mu\pi^i b_\mu^j\sigma + {1\over v}
   [\bar{\psi_L}\Phi\Lambda\psi_R+H.c.]\)^2 \nonumber \\ & & \qquad
  -4g_{ij} D^\mu\pi^i D_\mu\pi^j\( g_{lm} D^\nu \pi^l b_\nu^m
     \sigma+{1\over v}[\bar{\psi_L}\Phi\Lambda\psi_R+H.c.]\)
   \nonumber \\ & & \qquad
 + 2i\bar{\psi}_L[\Lambda\Lambda^{\dag},\Phi]\Phi^{\dag}\DS \psi_L
 -2i\bar{\psi}_L(\DS\Phi)\Lambda\Lambda^{\dag}\Phi^{\dag}\psi_L
 +2i\bar{\psi}_R\Lambda^{\dag}(\Phi^{\dag}\DS\Phi)\Lambda\psi_R
   \nonumber \\ & & \qquad
 -{1\over 6} F^3_{\mu\nu} G^{\mu\nu} + {1\over6}
   (K_{\mu\nu\; jk}^{\;\;\;\;\; i}\pi^k + 2R_{\mu\nu\; j}^{\;\;\;\;\; i}
     +2{\cal F}_{\mu\nu\; j}^{\;\;\;\; i})K^{\mu\nu\; j}_{\;\;\;\; il}
       \pi^l \nonumber \\ & & \qquad \left.
 -{2\over3}(\pe\cdot D_\nu\pe)\[ \FF^{\mu\nu}\cdot
   (\pe\times D_\mu\pe)+G^{\mu\nu}(\pe\times D_\mu\pe)^3\] \right\}.
\end{eqnarray}
Here
\begin{eqnarray}
 Z_g &=&  1-{13g^2\over 64\pi^2}\ln[\rho\mu^2/\mu_0^2], \nonumber \\
 Z_{g'} &=& 1+{81 g'^2\over192\pi^2} \ln[\rho\mu^2/\mu_0^2],
    \nonumber \\
 Z_v &=& 1-{\eta\mu^2/v^2\over 8\pi^2} +{\Tr \Lambda^{\dag}\Lambda
       \over 4\pi^2}\ln[\rho\mu^2/\mu_0^2],
    \nonumber \\
 Z_L &=& 1-{\ln[\mu_0^2/\rho\mu^2]\over 16\pi^2} \[
   \tr\Lambda\Lambda^{\dag}-{1\over2}\Lambda\Lambda^{\dag}\],
\nonumber \\
 Z_R &=& 1-{3\over2}\Lambda^{\dag}\Lambda{\ln[\mu_0^2/\rho\mu^2]
    \over 16\pi^2},
\nonumber \\
 Z_\Lambda &=& 1-{3\eta\mu^2/v^2\over 32\pi^2} +
   {\ln[\mu_0^2/\rho\mu^2]\over 8\pi^2}[\tr\Lambda\Lambda^{\dag}
   -{1\over2}\Lambda\Lambda^{\dag}] +
   {3\ln[\mu_0^2/\rho\mu^2]\over 64\pi^2} g'^2 Y_L\Lambda Y_R
    \Lambda^{-1},
\end{eqnarray}
are renormalization factors. For compactness of notation, the
terms involving the fermions $\psi$ are implicitly summed over
leptons and quarks, and $\Lambda$ embodies all the Yukawa couplings
in a large matrix:
\begin{equation}
 \Lambda =\(\begin{array}{cc} \lambda^{ab}_u[\hf+\tau_3]
    +\lambda^{ab}_d[\hf-\tau_3] & 0 \\  0 &
    \lambda_e^{ab} [\hf-\tau_3] \end{array}\), \qquad\quad
 \psi=\(\begin{array}{c}  q^a\\l^a \end{array} \),
\end{equation}
where $q^1, q^2, q^3$ are the three generation of colour triplet
quarks and $l^1, l^2, l^3$ are the three generations of colour
singlet leptons. The lower--case ``tr'' is a trace only over the
pauli matrices in (29), and $Tr \Lambda^{\dag}\Lambda =
 3\lambda^{\dag ab}_u\lambda_{u\; ab}+ 3\lambda^{\dag ab}_d
\lambda_{d\; ab}+\lambda^{\dag ab}_e\lambda_{e\; ab},$ with a sum
over $ab$. The factors of 3 are due to colour. $Y_L$ and $Y_R$
are also large matrices here. They are diagonal and their entries
correspond to the hypercharge assignments of the different fermions.
To be exact, from the covariant derivatives (9) and (10), we have
$Y_L(l^a_L)=-1, Y_L(q^a_L)=1/3, Y_R(l^a_R)=-2$, and
$Y_R(q^a_R)=$diag$(4/3,-2/3)$.
With this understanding, $\Phi$ in (27) is also a large matrix in
generation space. It is block diagonal, with each entry corresponding
to eq. (2).

The other undefined quantities in the above
equation are
\begin{eqnarray}
  S^\mu_{jI} &=& g_{ij}D^\mu\pi^l\times
    \left\{ \begin{array}{c} \epsilon^i_{\; 3l}+\delta^i_3\sigma_l
    +\pi^i  g_{lm}(\epsilon^m_{\;\; 3q}\pi^q+ \delta^m_3\sigma),
      \qquad I=0, \\
   \epsilon^i_{\; pl}-\delta^i_p\sigma_l+\pi^i  g_{lm}
   (\epsilon^m_{\;\; pq}\pi^q -\delta^m_p\sigma),\qquad I=p,
   \end{array} \right. \nonumber \\
   R_{\mu\nu\; j}^{\; \;\;\;\; i} &=& g_{jq}D_\mu\pi^i
      D_\nu\pi^q - (\mu\leftrightarrow\nu),
    \nonumber \\
   {\cal F}_{\mu\nu\; j}^{\;\;\;\;\; i} &=& \hf\epsilon^i_{\; mj}
       F^m_{\mu\nu}+\hf \epsilon^i_{\; 3j}G_{\mu\nu},
   \nonumber \\
   K_{\mu\nu\; jk}^{\;\;\;\;\; i} &=& \pi^i g_{lj}
      {\cal F}_{\mu\nu \; k}^{\;\;\;\;\; l} + O(\pe^2).
\end{eqnarray}
Also, $b_\mu^j = \hf A_\mu^j$ for $j=1,2$, and $b_\mu^3 =\hf
  A_\mu^3-\hf B_\mu^3$, and the sum over $p$ in the $S^2$ term of
eq. (27) is over $SU(2)_L$ gauge indices. The complete form of
the matrix $K_{\mu\nu}$ can be extracted from [2], but the
higher order terms are suppressed by inverse powers of $v$ and
are not phenomenologically interesting.

We have introduced $\rho$ and $\eta$ to parameterize the dependence
of the exact answer on the scheme chosen to regularize the divergent
integrals. For the double subtraction scheme that was used [5,17],
$\eta=2\ln 2$ and $\rho=\hf$. In our methodology, these may also
be taken to parametrize the uncertainty about the detailed way in
which the underlying theory enters to dampen the otherwise divergent
integrals. In this case we should properly use different parameters
for the different terms in our one--loop corrections. Of course, some
of the $\ln\mu^2$ terms should not be taken as finite in the sense
we have advocated -- these are the renormalizable terms that appear
regardless of whether we work with a linear or nonlinear sigma model.
The corrections to the gauge kinetic terms, for example, determine
the $\beta$--functions for the running of the gauge coupling
constants. We agree with previous results [6,18]. Similarly, the
$\ln\mu^2$ corrections to the Yukawa couplings determine their
$\beta$--functions. For the nonrenormalizable terms we identify the
cutoff $\mu$ with the physical mass of the Higgs (or the heaviest
$CP$--even Higgs in the two Higgs doublet model). The characteristic
scale $\mu_0$ is more problematic. Since these corrections arise
from diagrams containing the pseudo--goldstone bosons the most
natural choice is the W--boson mass. However, one has to be more
careful. To obtain the correct identification means evaluating terms
subleading to those found here since we were only interested in
the ultraviolet divergent terms. This is the reason for the artificial
infrared divergence for $\mu_0\ll M_H$ which should only happen for
special values of the external momentum. By examining the infrared
divergences more carefully and resuming the derivative expansion,
Cheyette and Gaillard
[6] have shown that in this limit the leading kinematic
factor for scattering
processes involving four external scalars is given by
 the replacement $\mu_0^2\rightarrow\partial^2$.

Our result (27)
agrees with those of Refs. [6] and [7] when it was possible
to compare, except for the
$(D_\mu\pi^i)^2[\bar{\psi}\Phi\Lambda\psi+H.c.]$
term on the sixth line for which we find the opposite sign to the
corresponding term in eq (4.23) of [7]. In ref. [19]
a similar computation
was performed for the terms involving both fermions and bosons that come
from mixed fermionic and bosonic loops (our $\Sigma$ terms). The
corrections are given in eq. (3.10a,b,e,f,h-j) of that paper. We find
complete agreement except for the only factor these authors did not
explicitly compute, eq. (3.10j), for which we find an answer three
times as big. However this term, which corrects the kinetic term for the
right--handed fermions (i.e. $Z_R$)
does not seem to contain much interesting physics. One further term
bears comment: the $S^2$ term on the third line of (27). This term
gives the scalar loop contribution to the $\rho$ parameter, already
found in a slightly different form for the nonlinear $\sigma$--model by
Cheyette [6], and it can be shown to agree with the standard result [20].

The first three lines of (27) contain terms appearing in the tree
Lagrangian, eq. (6), and may be renormalized by the addition of suitable
counterterms. For example, Costa and Liebrand [7] consider some of the
resulting
wave--function renormalizations, and so on. All the quadratically
divergent corrections are renormalizable so that physically measurable
effects are insensitive to the exact value of this cutoff, in agreement
with M. Veltman's screening theorem [21].

We would like to stress that our background field calculation gives
corrections manifestly invariant under $SU(2)_L\times U(1)_Y$
gauge transformations, as well as reparameterizations of the
scalar fields. We could fix to the unitary gauge and remove all
dependence on the pseudo--scalars, and in this form the corrections
to physical scattering processes may be more transparent. For example,
the first term on the fourth line of (27) is proportional to the
square of the tree level scalar kinetic term, so that in the unitary
gauge this term contains four gauge bosons and describes corrections
to such processes as $W_L W_L\rightarrow W_L W_L$. However, for the
energy range we shall be interested in (next section) it is more
convenient to think of the $\pe$ as the longitudinal components of
the massive gauge fields and use the equivalence theorem [22] to extract
the relevant cross--sections.

In section three we write out some of the terms in (27) for the
top/bottom fermion doublet. This gives an indication of how to
extract corrections for other processes.

\noindent {\bf Two Higgs Doublet Model.}

For the two
Higgs doublet model given by eq. (17), it is straightforward
to extend the previous results. We find $in$ $addition$ to the
results of (27) the following corrections:
\begin{eqnarray}
 \delta L &=& -{\mu^2\ln 2\over 16\pi^2} \[M_2+M_4\]
 \nonumber \\ & &
  -{\ln[\mu_0^2/\rho\mu^2]\over64\pi^2}\{ V_4
  - {1\over12}\FF_{\mu\nu}\cdot\FF^{\mu\nu}-{1\over12}
    G_{\mu\nu}G^{\mu\nu}-{1\over6}F_{\mu\nu}^3G^{\mu\nu}
  -{3\over2}{\cal S} \}
 \nonumber \\ & &
  -{\ln[\mu_0^2/\rho\mu^2]\over64\pi^2}\left\{ V_8 +
   2u\[8\tilde{V}+3V'-8\lambda_4'(\chi^{\dag}\phi)(\phi^{\dag}\chi)
    + 4v^2\lambda_4'(\chi^{\dag}\chi)\] \right.
 \nonumber \\ & &\qquad
   +2g_{ij}D_\mu\pi^iD^\mu\pi^j\[-2V'-6\tilde{V}+6\lambda_4'
     (\chi^{\dag}\phi)
(\phi^{\dag}\chi)-4v^2\lambda_4'(\chi^{\dag}\chi)\]
 \nonumber \\ & &\qquad
 +2D_\mu\pi^iD^\mu\pi^j
\[\lambda_4'[(\chi^{\dag}\phi_i)(\phi^{\dag}_j\chi)
    +H.c.]-{1\over2}
(\lambda_5-\lambda_6)[ (\phi^{\dag}_i\chi)(\phi^{\dag}_j
     \chi)+H.c.] \right \} .
 \nonumber \\ & & \qquad
\end{eqnarray}
Here, $M_2, M_4, V_4, V_8$ are corrections to the potential
terms. $M_2$ comes from loops involving the $\chi$ and gauge
fields only; $M_4$ and $V_8$ from loops containing the $\pe$ fields
only.
We have:
\begin{eqnarray}
 M_2      &=& +(3\lambda'+6g^2+2g'^2-4\lambda_4') \chi^{\dag}\chi
+(2\lambda_4+\lambda_5
     +\lambda_6)\phi^{\dag}\phi
 \nonumber \\
 M_4 &=&
   +8\lambda'_4 v^{-2}
     (\chi^{\dag}\phi)(\phi^{\dag}\chi)
  + 6v'/v  \[\lambda_5\cos\xi\Re(\phi^{\dag}\chi)+
      \lambda_6\sin\xi\Im(\phi^{\dag}\chi)\]
 \nonumber \\ & & \qquad
-2(\lambda_5-\lambda_6)
  v^{-2} \[ (\phi^{\dag}\chi)^2+H.c.\].
 \nonumber \\
 V_4 &=& \[{5\over2}\lambda'^2+3g^4+2g^2g'^2+g'^4\]
(\chi^{\dag}\chi)^2
 \nonumber \\ & &
+\[-3 \lambda'^2 v'^2 +(6g^4+2g'^4-4g^2g'^2)v^2
   + 4\lambda_4^2 v^2+ \lambda'(5\lambda_4+\lambda_5)v^2\]
      (\chi^{\dag}\chi)
   \nonumber \\ & &+\[
    -2\lambda'\lambda_4' +8g^2g'^2\]
       (\chi^{\dag}\phi)(\phi^{\dag}\chi)
 \nonumber \\
 v^4 V_8 &=& 20\lambda_4'^2(\chi^{\dag}\phi)^2(\phi^{\dag}\chi)^2
-16\lambda_4'^2v^2(\chi^{\dag}\phi)(\phi^{\dag}\chi)(\chi^{\dag}\chi)
   \nonumber \\ & &
 -\lambda_4'(\chi^{\dag}\phi)(\phi^{\dag}\chi)\[16V'+40\tilde{V}\]
  + v^2\lambda_4'(\chi^{\dag}\chi)\[8V'+16\tilde{V}\]
  \nonumber \\ & & + 3V'^2 + 20\tilde{V}^2
  \nonumber \\ & &
    -\lambda_4'(\lambda_5-
   \lambda_6)\[(\chi^{\dag}\phi_i)(\phi^{\dag}_j\chi)
       + H.c.\]\[ (\phi^{\dag}_k\chi)(\phi^{\dag}_l\chi)+H.c.\]
        g^{ik}g^{jl}
 \nonumber \\ & &
    +\lambda_4'^2\[(\chi^{\dag}\phi_i)(\phi^{\dag}_j\chi)
       + H.c.\]^2
 \nonumber \\
 u &=& g_{lm}D^\mu\pi^l b^m_\mu\sigma + {1\over v}
   \[\bar{\psi}_L\Phi\Lambda\psi_R + H.c.\]
 \nonumber \\
 {\cal S} &=& {\rm k.e.\; terms\; for\; scalars},
 \nonumber \\
 {\cal T} &=& i\bar{\psi}_L Z_L'\DS\psi_L+i\bar{\psi}_R Z_R'\DS\psi_R
     - v'[\bar{\psi}_L\ \Xi Z_\Lambda'\Lambda'\psi_R+ H.c.] .
\end{eqnarray}
We have defined $\lambda'_4=\lambda_4-{1\over2}\lambda_5-
{1\over2}\lambda_6$. We also have:
\begin{eqnarray}
 V' &=& -2\lambda_5 vv'\cos\xi\Re(\phi^{\dag}\chi)
      -2\lambda_6 vv'\sin\xi\Im(\phi^{\dag}\chi),
 \nonumber \\
 \tilde{V} &=& {1\over4}(\lambda_5-\lambda_6)
    \[(\phi^{\dag}\chi)^2+(\chi^{\dag}\phi)^2\],
\end{eqnarray}
$\Lambda'$ embodies all the Yukawa couplings of (18)
in a large matrix:
\begin{equation}
 \Lambda' =\(\begin{array}{cc} \lambda'^{ab}_u[\hf+\tau_3]
    +\lambda'^{ab}_d[\hf-\tau_3] & 0 \\  0 &
    \lambda_e'^{ab} [\hf-\tau_3] \end{array}\).
\end{equation}

The first two lines of (31) are renormalizable, it is only the
following lines which contain new terms. ${\cal S}$ contains the leading
Higgs mass dependent correction to the kinetic terms of the $\chi$.
${\cal T}$ are corrections from loops involving fermions and
$\chi$ fields, and they modify the kinetic terms and Yukawa couplings
of the fermions. The quantities $Z_L'$, $Z_R'$, $Z_\Lambda'$ which
give the precise one--loop shifts to the fermion kinetic and mass
terms are as in the nonmassive two Higgs doublet models and can be
easily deduced from the results of [2]. We do not give them here,
as they are not important to the physical processes we will
be interested in. (The shifts to the scalar kinetic energy will give
additional contributions to the $\rho$ parameter, but this is
already well known [14].)In addition, in contrast to (27), we have
normalized the gauge fields so that there are no overall factors
of gauge couplings outside the gauge kinetic terms.

The nonrenormalizable terms contain corrections with
multiple scalar fields, as well as couplings between fermions and
scalars. Our
interpretation of the scale $\mu$ will be the physical mass
of the $CP$--even Higgs we eliminated from the theory, i.e. $H^0$
in (15). To find the corrections to processes involving physical
particles one has to rewrite the sum of (27) and (31) by use of the
physical scalars of (14), and then to eliminate the ``eaten'' scalars
of (13) by going to the unitary gauge (or to use the equivalence
theorem for these in the appropriate energy range).

\vskip 0.3cm
\noindent {\large\bf 3. Physical Results.}

Amplitudes computed from our effective Lagrangians should give the
tree + leading Higgs mass dependent one--loop results. These are
the leading corrections for large enough Higgs mass, and for
illustrative purposes we consider some physical processes when
this is the case.

\noindent {\bf One Higgs Doublet Model.}

For the
strongly interacting standard model, one--loop corrections to the
Lagrangian were already computed for pure boson loops in [6,7]. In [6]
these corrections were
used to study the consequences for longitudinally
polarized gauge boson rescatterings. We focus here on corrections due
to loops with internal fermions and bosons, not computed for the
standard model before. These have two or more external fermions, and
any number of external pseudo--goldstone scalars and/or $SU(3)_c\times
SU(2)_L\times U(1)_Y$ Yang--Mills gauge fields. We kept only the leading
Higgs mass dependent ($M_H^2$ or $\ln M_H$) corrections.
These corrections contain only two external fermions,
and at most one external gauge boson. They may contain any number of
external scalars since we have taken the nonlinear limit in which
even the tree
level Yukawa couplings contain fermion interactions with an
arbitrary number of scalars. We expect that
loops containing more than two external
fermions or more
than one external gauge boson yield much smaller corrections
for large enough Higgs mass.

These mixed fermion/boson loop corrections
depend on the Yukawa couplings, or equivalently the fermion masses, so
we expect them to be largest for the top quark whose mass is at least
89 GeV [9,10] and is perturbatively constrained by the one--loop
$\rho$--parameter to be within the approximate range 125 GeV to
195 GeV in the standard model with a Higgs mass between 0.5TeV and 1TeV
[10]. ( Both the upper and lower
limits are lowered by as much as 40 GeV for a Higgs mass near
100 GeV.)
Taking our result of (27), ignoring $KM$--matrix mixing
angles and writing it out in full for the top/bottom doublet,
we have the following one loop correction to the tree--level
Lagrangian:
\begin{eqnarray}
\delta L &\ni& i{\ln[M_H/M_W]\over 32\pi^2 v^4}\left\{
  (m_t^2-m_b^2)\bar{\psi}_L [2\tau_3,\Phi]\Phi^{\dag}\DS\psi_L -
   (m_t^2+m_b^2)\bar{\psi}_L(\DS\Phi)\Phi^{\dag}\psi_L\right.
 \nonumber \\ & & \qquad\quad
 -(m_t^2-m_b^2)\bar{\psi}_L (\DS\Phi)(2\tau_3)\Phi^{\dag}\psi_L
  +{(m_t^2+m_b^2)^2\over 2}\bar{\psi}_R(\phi^{\dag}\DS\Phi)\psi_R
 \nonumber \\& & \qquad \quad\left.
 +{(m_t^2-m_b^2)\over2}\bar{\psi}_R\{2\tau_3,(\Phi^{\dag}\DS\Phi)\}
   \psi_R+ {(m_t-m_b)^2\over2}\bar{\psi}_R(2\tau_3)(\Phi^{\dag}
  \DS\Phi)(2\tau_3)\psi_R \right\}.\nonumber \\ & &
\end{eqnarray}
In this equations $M_H$ is meant to be taken as the mass of the
physical Higgs, $v$ is the Higgs $vev$,
and we have taken the $W$--boson mass for $\mu_0$. In addition, we
have returned to dimensionful scalar fields which are given by
\begin{equation}
 \Phi = \( \begin{array}{cc} \sqrt{v^2-\pe^2}+i\pi^3 &
  i\sqrt{2}\pi^+ \\ i\sqrt{2}\pi^- & \sqrt{v^2-\pi^2}-i\pi^3
    \end{array} \),
\end{equation}
where $\pi^\pm={1\over\sqrt{2}}(\pi^1\mp i\pi^2)$. In our unconventional
normalization for the pauli matrices, $2\tau_3=$diag$(1,-1)$. The
fermions are doublets
\begin{equation}
  \psi = \( \begin{array}{c} t \\ b \end{array} \).
\end{equation}

We are interested in
processes involving longitudinally polarized electroweak
gauge bosons. Here,
we can make use of the famous equivalence theorem [22]
which states that
physical amplitudes involving the longitudinal bosons are
related to unphysical
amplitudes involving the pseudo--goldstone bosons via
\begin{equation}
 A\[W^\pm_L(p),\ldots,Z^0_L(k),\ldots\] =
   A\[\pi^\pm(p),\ldots,\pi^3(k),\ldots\]+O\({M_W\over E}\).
\end{equation}
On the RHS the matrix element is to be evaluated in a gauge in which
the pseudo--goldstone
scalars still appear in the Lagrangian, and $E$ is
the energy of the process. This holds for any number of external states,
including those other than the longitudinal bosons. More importantly
it holds to all orders in the Higgs self--coupling $\lambda$, which is
important in the case of a strong $\lambda$. Although we could take
our tree+one--loop Lagrangian, fixed to the unitary gauge, and
calculate the actual gauge boson scattering amplitudes this turns
out to be more work than finding the corresponding amplitudes
involving the $\pe$. Using this procedure, our calculations should
faithfully reproduce the full answers in the energy range
\begin{equation}
    M_W \ll E \ll M_H.
\end{equation}
One can now see that equation (35) contains some one--loop corrections
to many processes, e.g. $\bar{t}t,\bar{b}b\rightarrow W_L^+ W_L^-$, or
$\bar{t}b\rightarrow W_L^- Z_L^0$. Since $\Phi$ contains any number of
$\pe$, the final states may contain any number of longitudinal bosons
(however each additional boson suppresses the amplitude by $v$).

There are other terms in (27) that contribute to such processes.
Consider $W_L^+ W_L^-$ which can proceed through an intermediate
photon or $Z^0$,
as well as bottom exchange. In the linear $\sigma$--model
(i.e. keeping the Higgs)
it also proceeds through an intermediate Higgs.
In our case the corresponding diagram is an elementary four point
vertex (the Higgs propagator is ``shrunk" to a point). We need to
examine all the one--loop corrections to these. In our analysis
here we ignore the finite fermion loops. One--loop corrections to, e.g.,
the $Z^0\pi^+\pi^-$ coupling are very small (as a case in point,
consider the term that gives the Higgs correction to the $\rho$
parameter). In fact, we find the leading correction apart from (35)
is a correction to the elementary four point vertex (the ``Higgs
exchange") as given on the sixth line of (27). Writing it out for the
top/bottom doublet with the same normalization as (35) we find
\begin{equation}
  \delta L \ni -{\ln[M_H/M_W]\over 4\pi^2 v^4} \left\{
    m_t \bar{t}t+ m_b\bar{b}b\right\} D_\mu\pi^+ D^\mu\pi^-.
\end{equation}
This correction is due to loops involving the pseudo--goldstone
scalars only, with two external scalars and an external fermion
bilinear $\bar{\psi}\psi$. The relevant tree level terms that we
must compare with are found from (6) to be
\def\AS{\not{\hspace{-0.045in}A}}
\def\BS{\not{\hspace{-0.050in}B}}
\begin{eqnarray}
  L_{tree} &=& {m_t\over v^2}\pi^+\pi^-\bar{t}t +
     {m_b\over v^2}\pi^+\pi^- \bar{b} b
   \nonumber \\ & &
 + i\sqrt{2}{m_t\over v}(\pi^+\bar{t}_R b_L-\pi^-\bar{b}_L t_R)
 + i\sqrt{2}{m_b\over v}(\pi^-\bar{b}_R t_L-\pi^+\bar{t}_L b_R)
   \nonumber \\ & &
 +{g\over 2}\bar{t}_L \AS^3 t_L+{g'\over 6}\bar{t}_L\BS t_L
   +{2g'\over3}\bar{t}_R\BS t_R + {g\over2}
     \bar{b}_L\AS^3 b_L + {g'\over6}\bar{b}_L\BS b_L -
    {g'\over 3}\bar{b}_R\BS b_R
 \nonumber \\ & &
  +{i\over2}(g A^3_\mu+g' B_\mu)\left\{\pi^- D^\mu\pi^+ -
     \pi^+ D^\mu \pi^-\right\} .
\end{eqnarray}
We have reinserted the gauge coupling constants $g$ and $g'$ for the
$SU(2)_L$ gauge fields $\AA_\mu$ and hypercharge gauge field
$B_\mu$, respectively, as compared to (6).

{}From the tree amplitudes of (35), (40), and (41) one can extract
the exact answer for the tree + one--loop amplitudes found from
the tree Lagrangian (up to subleading terms we have neglected).
Let us consider two cases within the context of the standard model.

\noindent {\bf The process }$\bar{t}t\rightarrow W_L^+ W_L^-$.
Neglecting $m_b$ in comparison to $m_t$ we find the relevant one--loop
corrections are
\begin{eqnarray}
  \delta L &\ni & im_t^2{\ln[M_H/M_W]\over 16\pi^2 v^4}\left\{
  \bar{t}_L\gamma^\mu t_L\(\pi^- D_\mu\pi^++\pi^+ D_\mu\pi^-\)
   \right. \nonumber \\ & & \qquad \qquad
 \left. +\bar{t}_R\gamma^\mu t_R \(\pi^+D_\mu\pi^- -\pi^- D_\mu\pi^+\)
  +2\pi^+\pi^-\bar{t}_L\DS t_L\right\}
   \nonumber \\ & &
 +\sqrt{2} m_t^2{\ln [M_H/M_W]\over 16\pi^2 v^3} \left\{
    \bar{b}_L\gamma^\mu t_L D_\mu\pi^- +\pi^-\bar{b}_L\DS t_L
    -\pi^+\bar{t}_L\DS b_L\right\} \nonumber \\ & &
 -m_t {\ln [M_H/M_W]\over 4\pi^2 v^4} D_\mu\pi^+ D^\mu\pi^-
  \bar{t}t .
\end{eqnarray}
These corrections are subdued by powers of $v$ compared to the tree
Lagrangian of (41) so we expect them to be largest for large top mass
and energy. For top
masses closer to the CDF limit than the perturbative
upper bound we may expect that at high enough energies the dominant
correction is from the last term in (42). In fact, H. Veltman
[8] has shown this to be the case by explicit computation of all
the leading amplitudes. Ref. [8] also shows that the dominant tree
contribution in this case is what corresponds to our four point
vertex, the first term in (38). To find the ratio of tree+one--loop to
tree amplitude we use $\partial^\mu\pi^+\partial_\mu\pi^-
\rightarrow -(p_+\cdot p_-)\pi^+\pi^- \rightarrow -\hf \hat{s}
\pi^+\pi^-$, where $p_\pm$ is the four--momentum of the outgoing
$\pi^\pm$ and $\hat{s}=(p_+ +p_-)^2$. In addition, following
Cheyette and Gaillard [6] we will use $\ln[M_H^2/M_W^2]\rightarrow
\ln[M_H^2/(-\hat{s})]$ for the correct kinematic factor for our
scattering process. Then we immediately find
\begin{equation}
   {A_{tree+one-loop}(\bar{t}t\rightarrow \pi^+\pi^-) \over
      A_{tree}(\bar{t}t\rightarrow \pi^+\pi^-) } \approx
   1 + {1\over 16\pi^2}{\hat{s}\over v^2}\ln[M_H^2/(-\hat{s})],
\end{equation}
in agreement with ref. [8].
For example, this can lead to about a 15\%
enhancement over the tree amplitude when $M_H\approx 2$TeV and
$\sqrt{\hat{s}}\approx 1$TeV. By the equivalence theorem this is
also an approximation to the correction $\bar{t}t\rightarrow
W_L^+ W_L^-$ so we may expect a 30\% enhancement in the elementary
cross--section with our values. Our result provides an important and
independent check of the result of ref. [8].

For $m_t$ closer to $v$
the other tree diagrams from (42) must be more
carefully considered.

{\bf The process $W_L^+ W_L^- \rightarrow \bar{\Psi}\Psi$} for
heavy fermion pair production. Since mass splittings between a
fermion doublet contribute to the one--loop $\rho$--parameter
we consider an almost degenerate fermion doublet $(\Psi,\Xi)$
with average mass M. In this case the leading corrections similar
to (35) and (40) for this process are:
\begin{eqnarray}
\delta L \ni & & -iM^2 {\ln [M_H/M_W]\over 8\pi^2 v^4}
  (\pi^-D_\mu\pi^+ -\pi^+ D_\mu\pi^-)\bar{\Psi}\gamma^\mu\Psi
  \nonumber \\ & &
 -\sqrt{2}M^2{\ln[M_H/M_W]\over 16\pi^2 v^3}
    \[\bar{\Psi} (\DS\pi^+)\Xi+H.c.\]
 -M{\ln[M_H/M_W]\over  4\pi^2 v^4} D_\mu\pi^+ D^\mu\pi^-
    \bar{\Psi}\Psi. \nonumber \\ &&
\end{eqnarray}
We must compare this with
the corresponding tree contributions, a minor
modification of (41), and also production via quark--antiquark
and gluon fusion. This process was examined by Costa and Liebrand [7]
who only considered corrections from the last term in (44). For
$M^2 \sim  \hat{s}$ the other terms are also important and a
more complete analysis is necessary. For heavy quark production,
the QCD background at the LHC or SSC will be enormous [23] so our
extra terms are do not alter the conclusion of [7] that
one--loop corrections give only an infinitesimal contribution to
the total rate. For equally heavy leptons, the QCD background
is much smaller so these corrections may be observable
(if they are detectable at all, since they will decay into a sea
of $W^+ W^-$).

{\bf Two Higgs Doublet Model.}

Our corrections (31) in the case of a strongly interacting two Higgs
doublet model are entirely new. Of course, there are many more
parameters than in the minimal standard model, and we have only
studied the case when one of the $CP$--even physical scalars is very
massive. The leading fermion mass dependent corrections, as mentioned
previously, will be presented
elsewhere. However, we are able to extract
what should be the leading corrections at high enough energy. This
would be physically relevant if some evidence for at least an extra
Higgs doublet is found (say $CP$--odd scalars like $H^\pm$) but
the scalar sector is strongly interacting.

It is well known [1] that tree level amplitudes in the minimal
standard model with very large Higgs self--coupling violate
unitarity at $\sqrt{s_{c}}$=1.7 TeV.
The tree level unitarity
bound can be simply found by considering the nonlinear $\sigma$--model
given by (4). This Lagrangian gives, using the equivalence theorem,
the leading two body to two body scattering amplitudes for processes
involving longitudinally polarized $W$s and $Z$s. In particular,
the $J=0$ partial wave amplitude $a_0$ for the process
$W_L^+ W_L^- \rightarrow Z_L Z_L$ grows with energy to lowest order as
\begin{equation}
   |a_0(W_L W_L\rightarrow Z_L Z_L)| \approx {s\over 16\pi v^2}
     = {s\over (1.7TeV)^2}
\end{equation}
for $\sqrt{s}$ much less than the $perturbative$ Higgs mass
$M_H=350\sqrt{\lambda}$GeV (see eq. (1)).
Demanding $|a_0|<1$ gives the approximate bound on $\sqrt{s}$ for the
scale at which new physics, for example the
physical Higgs, should appear to make the full theory unitary.

One loop corrections to such unitarity bounds (for the one
Higgs doublet case) have been considered in [4].
One finds that low energy unitarity bounds can be made
significantly stricter. While one should be careful in believing
perturbative results in the strongly interacting case, it is important
to note that such one--loop calculations support the belief [1]
that a Higgs or new physics should appear at or before SSC energies.

Since there is no $a$ $priori$ reason to belive in the minimal
standard model it is important to investigate other models to
understand what we may be able to detect at the next generation
of colliders. There are two useful limits: $\tan\beta=v'/v$ small
and $\tan\beta$ large. These limits correspond to the $\pe$ fields
of the strongly interacting scalar doublet being purely the
eaten bosons (the longitudinal gauge bosons) or being purely the
CP--odd physical Higgs, respectively. The relevant four pion
tree level vertex is in both cases given by (4).
For small $\tan\beta$ the $vev$ $v$ is fixed from experiment by
$1/v^2=\sqrt{2}G_F$ and the
the process\footnote{
Recall that the eaten bosons (or equivalently for us the longitudinal
gauge bosons) are given by (13). We use $z=\sqrt{2} w^0$. The
field $A^0$ is given by (14).}
$w^+ w^-\rightarrow z z$ gives the same tree level unitarity bound as
from (45), namely $\sqrt{s_c} <$1.7TeV. For large $\tan\beta$
the process $H^+ H^-\rightarrow A^0 A^0$ gives the same partial wave
amplitude as (45), but now $1/(v^2 + v'^2)=\sqrt{2}G_F$ fixes
$v^2 \approx v'^2/(tan\beta)^2$ and we get the bound
$\sqrt{s_c} < 1.7$TeV$/\tan\beta$. For $\tan\beta=2$ this gives
the scale of unitarity violation as just 875 GeV.

A more careful tree level analysis of unitarity in the
two Higgs doublet model has been carried out in [11]. One finds that
\begin{equation}
   \sqrt{s_c} = {1.7 TeV\over \sqrt{1+(\tan\beta)^2}},
\end{equation}
for the heavy Higgs
limit we assumed (see eq. (15)).
This result comes from
considering all the possible two body to two body scatterings
with neutral initial and final states, and finding the amplitude
$a_0$ with the largest absolute value.
Therefore, the energy at which
new phenomena may appear in the strongly interacting two Higgs
doublet model may be significantly lower than in the standard
model.

Since we have explicitly calculated the relevant leading terms
we can easily determine one loop corrections to these
unitarity bounds.
We consider only $w^+ w^-\rightarrow zz$ for small $\tan\beta$
and $H^+ H^-\rightarrow A^0 A^0$ for large $\tan\beta$.
Since at tree level these channels reproduce the exact result
for a wide range of $\tan\beta$,
they will give a
good indication of how one loop corrections change such unitarity
bounds.

The relevant tree+one--loop Lagrangian for large $\lambda_1$ is
given by (27) and (31).
We will probe the region $M_W^2 \ll s \ll 2\lambda_1 v^2$ so we
need only consider the derivative terms. Keeping the necessary
terms to lowest
order in the scalar fields and ignoring renormalizable corrections,
we find that the relevant terms are
\def\KE{ \delta_{ij} D_\mu\pi^ i D_{\mu} \pi^j}
\def\KK{ \delta_{ij} D_\mu\pi^ i D_{\nu} \pi^j}
\begin{eqnarray}
  L \ni & &  {1\over2} v^2 g_{ij}D_\mu\pi^i D^\mu\pi^j \nonumber \\
   & & \quad
+{\ln[M_H^2/\rho\partial^2]\over 64\pi^2}
     \[{2\over3} (\KE)^2+{4\over3}(\KK)^2 \right.
   \nonumber \\ & &\quad
  + 2\KE\(-2V'-6\tilde{V}+6\lambda_4'(\chi{\dag}\phi)(\phi^{\dag}\chi)
   -4v^2\lambda_4'(\chi^{\dag}\chi)\)
   \nonumber \\ & & \quad \left.
 + 2D_\mu\pi^i D^\mu\pi^j\(\lambda_4'[(\chi^{\dag}\phi_i)
    (\phi^{\dag}_j\chi) +H.c.]-{\lambda_5-\lambda_6\over2}
        [(\phi^{\dag}_i\chi)(\phi^{\dag}_j\chi)+H.c.]\)\].
\nonumber \\ & &
\end{eqnarray}
We have inserted the correct factor in the logarithm [6] for the
scattering process and substituted the perturbative Higgs mass
(of $H^0$) for $\mu$;  $\rho$ is undetermined in our approach
and since we do not have the next to leading one loop corrections
the best we can do is
assume it is of $O(1)$. In what follows we drop it.
In spite of this
uncertainty the above prescription should give a good indication
of the amount by which the one loop corrections modify the tree
level result.

{\bf Small $\tan\beta$.}
Writing out the leading terms in (47) that are important for $w^+ w^-
\rightarrow zz$,
we get
\begin{eqnarray}
 L\ni & & {1\over v^2 }
    \[ w^+ z (D_\mu w^-) (D^\mu z)+ H.c.\] \nonumber \\ & &
   + {\ln[M_H^2/     \partial^2]\over 64\pi^2}
         {8\over3 v^4}(D_\mu w^+)(D^\mu w^-)
         (D_\nu z)(D^\nu z)
   \nonumber \\ & & \qquad
   + {\ln[M_H^2/     \partial^2 ]\over 64\pi^2}
         {16\over3 v^4}(D_\mu w^+)(D_\nu w^-)
         (D^\mu z)(D^\nu z).
\end{eqnarray}
 Note that the one--loop corrections in the last two lines of
(47) do not contribute in this limit.

Eq. (48) tells us that the leading corrections are as in the
standard model for this process [6], and one--loop unitarity bounds
have been well studied [4,24].
Ignoring external masses,
the amplitude for $w^+w^-
\rightarrow z z$ is
\begin{equation}
  A(\hat{s},\hat{t},\hat{u})
 = {1\over v^2} \[s+{1\over16\pi^2 v^2}\[{s^2\over2}\ln[M_H^2
            /(-\hat{s})]+{2\hat{t}^2+\hat{s}\hat{t}\over6}
    \ln[M_H^2/(-\hat{t})]+{2\hat{u}^2+\hat{u}\hat{s}\over6}
    \ln[M_H^2/(-\hat{u})] \]\].
\end{equation}
The $J=0$ partial wave is easily found (see for example ref. [1])
by using $\hat{s}+\hat{t}+\hat{u}=0$ and integrating over $\hat{t}$.
We find for the real part
\begin{equation}
  a_0(\hat{s}) = a_0^t\[1+{a_0^t\over\pi}\[{73\over36}\ln[M_H^2/
    s]-{1\over108} \]\],
\end{equation}
where $a^t_0$ is the tree result    of (45), which does not depend
on the parameter $M_H$.
One finds, for example, that unitarity is saturated at $\sqrt{s_c}
=$1740 GeV when $M_H=1750$ GeV.

{\bf Large $\tan\beta$.} This case is much more interesting. It
is easy to check that the relevant Lagrangian for $H^+ H^-\rightarrow
A^0A^0$ is similar to (48). We have
\begin{eqnarray}
 L\ni & & {(\tan\beta)^2\over v'^2 }
    \[ H^+ A^0 (D_\mu H^-) (D^\mu A^0)+ H.c.\] \nonumber \\ & &
   + {\ln[M_H^2/     \partial^2]\over 64\pi^2}
         {8(\tan\beta)^4\over3 v'^4}(D_\mu H^+)(D^\mu H^-)
         (D_\nu A^0)(D^\nu A^0)
   \nonumber \\ & & \qquad
   + {\ln[M_H^2/     \partial^2 ]\over 64\pi^2}
         {16(\tan\beta)^4\over3 v'^4}(D_\mu H^+)(D_\nu H^-)
         (D^\mu A^0)( D^\nu A^0).
\end{eqnarray}
The partial wave amplitude for our process is then again given
by (50), with
\begin{equation}
  a_0^t(\hat{s}) = {(\tan\beta)^2\hat{s}\over16\pi v'^2}
= {(\tan\beta)^2\hat{s}\over (1.7TeV)^2}.
\end{equation}

The perturbative tree
level heavy Higgs mass grows with $\sqrt{\lambda_1}
/\tan\beta$  in this limit. To study the large mass limit we may study
either the limit $\lambda_1$ large and fixed or $M_H$ large and fixed.
In both cases large means $M_H$ is much bigger than $M_W$,
$M_{top}$ and $\sqrt{s}$.
To get an idea of $\sqrt{s_c}$ in this case we computed the scale
at which (50) is $1$ for two cases: (i) $M_H=1750$ GeV and (ii) $M_H=
1750$ GeV$/ \tan\beta$. For the last case, the tree+one--loop
unitarity bound is approximately the same as the tree level
bound, 1750 GeV$/(\tan\beta)$ for the values of $\tan\beta$
we looked at ($2< \tan\beta<10$). In fact the bound is a little
higher, since at $M_H^2= s= 1750GeV/(\tan\beta)$
   the logarithm in (50) vanishes and the remaining one--loop term
helps to unitarize the theory. However, since we have dropped
subleading one--loop corrections, as well as ignoring the
imaginary part of the partial wave, all we can say in this case
is that there is no significant shift from the tree level
unitarity result.

For case (i) with fixed $M_H$ the unitarity bound drops
faster at large $\tan\beta$ with
$\tan\beta$ than the tree level result.
For example, at $\tan\beta=2$ we get $\sqrt{s_c}=666$ GeV and at
$\tan\beta=10$ we get $\sqrt{s_c}=110$ GeV. This is to be
contrasted with the tree level results of $\sqrt{s_c}=875$GeV at
$\tan\beta=2$ and $\sqrt{s_c}=175$GeV at $\tan\beta=10$ in the
large $\tan\beta$ limit.
In our figure we plot
the tree+one loop  $\tan\beta=0$ unitarity bound
as well as the large $\tan\beta$ limit results for $2<\tan\beta<10$.
For extremely high values of $\tan\beta$ our results are not to be
trusted since $\sqrt{s_c}$ is driven near $M_W$, $M_{top}$ and our
approximations are not good.
For $\tan\beta$ in the approximate range $0.5$
to 2, one needs a more careful analysis starting with equation (47).
In this case it is expected that the last two terms in (47)
will be important, since they contribute for arbitrary $\tan\beta$
to any of the neutral channels.

In summary, our results indicate that in the case of a
two Higgs doublet model  either  signs of the possible
strong interactions of the scalars or the scalars themselves
may turn up at relatively low energies. At $\tan\beta=4$, this
scale can be as low as 300 GeV. In addition, if one
carries out a more detailed analysis keeping the subleading terms
and the imaginary parts (as well as perhaps higher order corrections
and top mass dependent one--loop corrections)
it would be possible to gain information about  $M_H$
as a function of $\tan\beta$
by studying cross-sections at future experiments.
For example,
the scale of unitarity violation given by (50) depends on $M_H$:
for $\tan\beta=3$ we have $\sqrt{s_c}=500,415,380$GeV
at $M_H=750,2000,4000$GeV, respectively.
This simple example demonstrates the interesting
interplay between unitarity, $\tan\beta$ and the Higgs masses in
the two Higgs doublet model.
One further analysis
which would be useful is to take all the perturbative physical
scalar masses to infinity, and then study unitarity violations.
We leave these as  future projects.

\vskip 0.30cm
\noindent{\large\bf Acknowledgements.}

I would like to thank Francesca Borzumati and Mary K. Gaillard
for very valuable discussions and advice. I also thank M. K. Gaillard
for reading the manuscript.

\newpage
\noindent{\large\bf References.}
\begin{enumerate}
\item M. S. Chanowitz, {\it Ann. Rev. Nucl. Part. Sci.} {\bf 38:}
  323 (1988) and Berkeley preprint LBL--28110 and references therein.
\item V. Jain, ``The Strongly Interacting Standard Model at One--Loop,"
 Annecy preprint LAPP--Th--297/90.
\item A. C. Longhitano, {\it Phys. Rev.} {\bf D22:} 1166 (1980)
 and {\it Nucl. Phys.} {\bf B188:} 118 (1981); \\
T. Appelquist and C. Bernard, {\it Phys. Rev.} {\bf D22:}
 200 (1980), {\bf D23:} 425 (1981).
\item M. Veltman and F. Yndurain {\it Nucl. Phys.} {\bf B325:}
 1 (1989); \\ S. Dawson and S. Willenbrok, {\it Phys. Rev. Lett.}
 {\bf 62:} 1232 (1989).
\item M. K. Gaillard, {\it Nucl. Phys. } {\bf B268:} 669 (1986).
\item O. Cheyette, Berkeley preprint LBL--23247 and
 {\it Nucl. Phys.} {\bf B297:} 183 (1988);\\
O. Cheyette and M. K. Gaillard, {\it Phys. Lett.} {\bf B197:}
 205 (1987).
\item K. Costa and F. Liebrand, {\it Phys. Rev.} {\bf D40:}
 2014 (1989).
\item H. Veltman, {\it Phys. Rev.} {\bf D43:} 2236--2258 (1991).
\item CDF Collaboration, F. Abe et. al., {\it Phys. Rev. Lett.}
  {\bf 64:} 142 (1990).
\item G. Altarelli, CERN--TH--6206/91.
\item R. Casalbouni, D. Dominici, R. Gatto, and C. Giunti,
 {\it Phys. Lett.} {\bf B178:} 235 (1986) and
 {\it Nucl. Phys.} {\bf B299:} 117 (1988).
\item See, for example, M. K. Gaillard, ``Effective Nonrenormalizable
 Theories at One Loop," in {\it Particle Physics, Carg\`ese 1987},
 ed. M. L\'evy et. al., Plenum Press, New York (1988), p. 189.
\item S. Glashow, J. Illiopoulos, and L. Maiani, {\it Phys. Rev.}
 {\bf D2:} 1285 (1970).
\item  J. Gunion, H. Haber, G. Kane, and S. Dawson,
``The Higgs Hunter's Guide".
\item S. Honerkamp, {\it Nucl. Phys.} {\bf 36:} 130 (1973); \\
 L. Alverez--Gaum\'e, D. Z. Freedman and S. Mukhi,
{\it Annals of Physics} {\bf 134:} 85 (1981); \\
 S. Mukhi, {\it Nucl. Phys.} {\bf B264:} 640 (1986).
\item J. Burton, M. K. Gaillard, and V. Jain, {\it Phys. Rev. }
 {\bf D41:} 3118 (1990).
\item P. Bin\'etruy and M. K. Gaillard, {\it Nucl. Phys. }
 {\bf B312:} 341 (1989).
\item T.-P. Cheng and L.-F. Li, ``Gauge Theory of Elementary Particle
Physics,'' {\it Oxford University Press} (1984).
\item G. Cveti\u{c} and R. K\"ogerler, {\it Nucl. Phys. } {\bf B328:}
342 (1989).
\item A. C. Longhitano, Ref. 2;\\ J. Van Der Bij and M. Veltman,
{\it Nucl. Phys.} {\bf B231:} 205 (1984).
\item M. Veltman, {\it Acta Phys. Pol.} {\bf B8:} 457 (1977);\\
  M. Veltman, {\it Nucl. Phys. } {\bf B123:} 89 (1977); \\
  M. S. Chanowitz, M. A. Furman and I. Hinchliffe, {\it Nucl. Phys.}
  {\bf B153:} 402 (1979).
\item J. Cornwall, D. Levin and G. Tiktopoulos, {\it Phys. Rev.}
  {\bf D10:} 1145 (1974); \\ C. Vayonakis, {\it Lett. Nuovo Cimento}
  {\bf 17:} 383 (1976); \\ B. Lee, C. Quigg and H. Thacker,
  {\it Phys. Rev.} {\bf D16:} 1519 (1977); \\ M. Chanowitz and M. K.
Gaillard, {\it Nucl. Phys.} {\bf B261:} 379 (1985);\\
 G. Gounaris, R. K\"ogerler and H. Neufeld, {\it Phys. Rev.}
{\bf D34:} 3257 (1986); \\ H. Veltman, {\it Phys. Rev.} {\bf D41:}
2294 (1990).
\item E. Eichten, I. Hinchliffe, K. Lane and C. Quigg,
 {\it Rev. Mod. Phys. } {\bf 56:} 579 (1984) and {\bf 58:}
 1065(E) (1986).
\item L. Durand, J. M. Johnson and P. N. Maher, Madison
 preprint MAD/TH/90-6.
\end{enumerate}
\end{document}